\documentclass[preprint,aps,showpacs]{revtex4}

\usepackage{graphicx}
\usepackage{dcolumn}
\usepackage{bm}
\usepackage{color}
\usepackage{upgreek}
\usepackage{amsmath}

\begin{document}


\title{Atomistic Green's Function Method supported by \textit{ab initio} Calculations : Application to phonon transport in ZnO and ZnS}

\author{M.~Bachmann} \author{M.~Czerner} \author{S.Edalati-Boostan} \author{C.~Heiliger}
\affiliation{I. Physikalisches Institut, Justus-Liebig-Universit\"at Giessen, Germany}

\date{\today}

\begin{abstract}
We present an approach to calculate ballistic phonon transport that combines the atomistic Green's function (AGF) method with \textit{ab initio} results. For the inter atomic potential we use the harmonic approach. The equilibrium positions of the atoms and the inter atomic force constants(ifcs) are calculated using the ABINIT program package \cite{Gonze2005}, which is based on density functional theory. Therefore, the presented approach is parameter free. From the Green's function of the system we determine the density of states as well as the transmission function. The thermal conductance is obtained within the linear response regime. We apply this approach to bulk ZnO and bulk ZnS. Transmission functions for different transport direction for each material are presented. A comparison of the transmission function shows, that a ZnO/ZnS interface could be a promising phonon blocker. Adding such interfaces in ZnO or ZnS based thermoelectric devices could therefore increase the figure of merit. 
\end{abstract}


\maketitle

\section{Introduction}

For thermoelectric devices the figure of merit ZT is the most important characteristic number since it is directly connected to the theoretical maximum achievable efficiency of the device. The figure of merit is defined as
\begin{equation}
ZT=\frac{\sigma S^2 T}{\kappa_e + \kappa_{ph}}
\label{eqn:zt}
\end{equation} 

where $\sigma$ is the electric conductivity, S is the Seebeck coefficient, T is the temperature, $\kappa_e$ is the electric part of the thermal conductivity and $\kappa_{ph}$ is the phonon part of the thermal conductivity. 
In 2001 Venkatasubramanian \textit{et al.}\cite{Venkat2001} showed that the figure of merit in thermoelectric devices which are based on super lattice structures can be rather large. The large figure of merit is explained by a strong decrease in the phonon heat conductivity due to additional scattering of phonons at the interfaces and simultaneously an almost uneffected electronic transport across these interfaces. Venkatasubramanian \textit{et al.} investigated super lattice structures that are based on $Bi_2Te_3/Sb_2Te_3$.
Recent experimental results show that also ZnO/ZnS based super lattice structures and micro structures are a promising candidate as thermoelectric materials \cite{Homm2010a,Homm2010b}. It can be grown as sputtered thin films as well as fully epitaxially using molecular beam epitaxy. The material parameters can be tuned by doping. 
In this paper we introduce a scheme that combines the flexibility of the atomistic Green's function (AGF) method and the \textit{ab initio} character of DFT to calculate ballistic phonon transport. We use this scheme to investigate the thermal conductance of pure ZnO and pure ZnS. ZnO has wurtzite structure while ZnS can occur in wurtzite structure and zincblende structure. We show that ZnO/ZnS interfaces are promising phonon blockers regardless of the ZnS structure.
ZnO has a band gap of 3.37eV and shows an intrinsic n-doping, which is related to intrinsic Zn interstitials \cite{Look1999}. The space group of wurtzite structure is P6$_3mc$. Since ZnO in wurtzite structure has four atoms in the unit cell the phonon-dispersion relation has 12 branches. At the $\Gamma$ point the 12 branches divide up into 2A1+2B1+2E1+2E2, where both E1 and both E2 modes are double degenerated. In principle, the phonon dispersion relation can be measured by means of inelastic neutron-scattering (ISN) or for the Raman active modes with Raman scattering. For ZnO several experimental data with both methods have been reported\cite{Damen1966,Hewat1970,Thoma1974,Serrano2003}.  
ZnS has a slightly higher band gap than ZnO of 3.8eV. Chen \cite{Cheng2009} \textit{et al.} reported experimental and theoretical investigations of ZnS phonons in wurtzite and zincblende structure. They observe a good agreement between \textit{ab initio} calculations and measurements. 
AGF is used in the literature to calculate ballistic phonon transport \cite{Zhang2007a,Zhang2007b,Mingo2003,Hopkins2009} and results show good agreement with experiment.
Other models that are used to calculate phonon transport are the acoustic mismatch model (AMM), which is only valid for the long wavelength phonons and the diffuse mismatch model (DMM), which assumes diffuse scattering at the interfaces \cite{Swa1989}. Both methods do not take into account the structure of the interface and give therefore only an approximative description of phonon transport.

\section{Method}
The whole calculation divides up into two different tasks. The first task is the calculation of the equilibrium positions of the atoms and the determination of the interatomic force constants (ifcs). In the second task the AGF method is used to calculate the transmission function and the thermal conductance, whereas the ifcs from the first task are used to describe the interatomic potential. With this approach only the harmonic part of the potential is considered, therefore anharmonic effects are not described. In principle anharmonic effects can be incorporated in this approach by introducing additional self energies in the AGF method.  

\subsection{Calculation of the Interatomic Force Constants}
In order to calculate the ifcs we use the code package ABINIT \cite{Gonze2005}. This code is based on density functional theory and can be used with a pseudopotential method and a plane-wave expansion. For the exchange correlation potential, we use the local density approximation(LDA). The pseudopotentials are represented in the Troullier-Martins scheme \cite{Troullier1991}. The code uses a perturbation method to calculate the dynamical matrix on a discrete grid in the Brillouin zone. The use of a perturbation method avoids the use of super cells. Then the ifcs can be obtained by a discrete inverse Fourier transformation of the discrete dynamical matrix. Knowing the ifcs the dynamical matrix can be calculated in the whole Brillouin zone. Furthermore, also based on the same perturbation method the code is able to calculate the dielectric permittivity tensor and the born effective charges.
Knowing the dielectric permittivity and the born effective charges the ifcs can be decomposed in a long range part which describes the dipole-dipole interaction and a short range part which describes the electronic contribution to the interaction. For details see References \cite{Gonze1997a} and \cite{Gonze1997b}.

\subsection{Transport Calculation using Atomistic Green's Function Method}
A good overview of the AGF method applied to phonons is given in Reference \cite{Zhang2007a}. Here we will summarize the basic features and show how to connect these method to the ifcs obtained from the \textit{ab initio} calculation. First, we define the interaction of the atoms in the system and set up the harmonic matrix H. In the literature these interactions are often described by parameterized potentials\cite{Zhang2007b,Hopkins2009}. We use the ifcs obtained from \textit{ab inito} calculations to describe the interaction between the atoms. The interaction between atom $\mu$ and atom $\eta$ is represented by a 3x3 matrix.

\begin{equation}
H_{\mu_i\eta_j}=\frac{1}{\sqrt{M_{\mu}M_{\eta}}}k_{\mu_i\eta_j}
\label{eqn:harmonic}
\end{equation}

$M_{\mu}$ and  $M_{\eta}$ are the masses of the atoms and $k_{\mu_i\eta_j}$ is the force constant that describes the impact on atom $\mu$ in direction i if atom $\eta$ is dislocated in direction j and vice versa. If we describe a structure that has periodicity in all 3 dimensions we can built the dynamical matrix which is the Fourier transform of the harmonic matrix. 
\begin{equation}
D(\vec{q})=\sum_{\vec{R}}H(\vec{R}) e^{-i\vec{q}\vec{R}}
\label{eqn:dynamical}
\end{equation}

$\vec{R}$ is a lattice vector in real space and $H(\vec{R})$ contains all interactions between cells that are connected by $\vec{R}$. The phonon dispersion relation can be obtained by the square root of the eigenvalues of the dynamical matrix. The density of states can be calculated using the Green's function of the system.

\begin{equation}
G(w,\vec{q})=(\omega^2I-D(\vec{q}))^{-1}
\label{eqn:Green}
\end{equation}

The Green's function and the density of states are connected by an integration over the Brillouin zone

\begin{equation}
n(\omega)= \frac{\mathrm{i}\omega }{\pi}\int_{BZ} \left(G(\omega,\vec{q})-G(\omega,\vec{q})^{\dagger}\right) d\vec{q}
\label{eqn:GDos}
\end{equation}

For transport calculations we consider a system that is coupled to two semi-infinite contacts. The considered system is in general not periodic in transport direction.   
Therefore, we divide our system into layers of atoms that are perpendicular to the transport direction. This allows us to distinguish between interlayer and intralayer interactions. The layers are still periodic perpendicular to the transport direction. Consequently we can Fourier transform the interlayer and intralayer interactions along these directions. The periodicity in-plane is described by a 2-dimensional lattice with the lattice vectors ${\vec{R}^{g}_p}$, where g is the layer index. The Fourier transform of the interlayer and intralayer interactions are given by

\begin{equation}
H_g(\vec{q}_p)=\sum_{\vec{R}^{g}_p}H(\vec{R}^{g}_p) e^{-i\vec{q}_p\vec{R}^{g}_p}  \text{    (intralayer)}
\label{eqn:Hp}
\end{equation}

\begin{equation}
T_{g\bar{g}}(\vec{q}_p)=\sum_{\vec{R}_o}H(\vec{R}_o) e^{-i\vec{q}_p\vec{R}_o}  \text{    (interlayer)}
\label{eqn:Hp2}
\end{equation}

$\vec{R}_o$ are the vectors that connect cells of neighbouring layers $g\bar{g}$. With this 2-dimensional Fourier transformation we introduce a new Vector $\vec{q}_p$, a 2-dimensional unit cell, and the corresponding 2-dimensional Brillouin zone. 

The overall matrix describing the system has then the following form. 

\begin{equation}
H(\vec{q}_p)=\begin{pmatrix} ... & T_{g-1,g} & 0 & 0 \\ T_{g,g-1} & H_{g} & T_{g,g+1} & 0 \\ 0 & T_{g+1,g} & H_{g+1} & T_{g+1,g+2} \\ 0 & 0 & T_{g+2,g+1}& ...  \end{pmatrix} 
\label{eqn:GDos2}
\end{equation}

$H_{g}$ is the Fourier transformed intralayer interaction of layer g. $T_{g,g+1}$ is the Fourier transformed interlayer interaction of layer g with g+1. In a typical calculation we take into account more than next nearest layer interaction.

This matrix is infinite. The infinite matrix can be reduced to a finite matrix by using the concept of self-energies. 
The overall Green's function of the system is given by

\begin{equation}
G(w,\vec{q}_p)=(\omega^2I-H-\Sigma_L-\Sigma_R)^{-1}
\label{eqn:GDos3}
\end{equation}

where H is a finite submaxtrix of H$(\vec{q}_p)$ and $\Sigma_L$ and $\Sigma_R$ are the self energies for the left and right contact, respectively \cite{Zhang2007a}. The self energies can be calculated using

\begin{align}
\Sigma_L=T_Lg_lT_L^{\dagger}\\
\Sigma_R=T_Rg_rT_R^{\dagger}\nonumber
\label{eqn:Sigma}
\end{align}

where T$_L$ and T$_R$ are the interlayer interactions which connect the middle region to the left and the right semi-infinite leads. $g_l$ and $g_r$ are the surface Green's functions of the left and right contact. We calculate these surface Green's functions using decimation techniques \cite{Guinea1983}.

From the self energies the broadening matrices can be obtained using

\begin{equation}
\Gamma_{L,R}=i(\Sigma_{L,R}-\Sigma_{L,R}^{\dagger})
\label{eqn:Tau}
\end{equation}

Consequently the transmission function of the system can be calculated using

\begin{equation}
t(\omega,q_p)=Tr[\Gamma_LG\Gamma_RG^{\dagger}]
\label{eqn:Trans}
\end{equation}

The average transmission function per unit cell is given by an integral over the 2-dimensional Brillouin zone

\begin{equation}
t(\omega)=\frac{1}{(2\pi)^2}\int_{BZ} t(\omega,q_p)dq_{p}
\label{eqn:Transmission}
\end{equation}

The unit of this quantity is transmission per area.

Knowing the average transmission one can calculate the total energy flux per unit area J in the linear response

\begin{equation}
J=\frac{1}{2\pi}\int_0^\infty \hbar \: \omega \: t(\omega) \frac{\partial f(\omega,T)}{\partial T}  d\omega  \Delta T
\label{eqn:J}
\end{equation}

where $f(\omega,T)$ is the occupation function for the phonons.

The conductance per unit area is defined as 

\begin{equation}
\kappa=\frac{J}{\Delta T}=\frac{1}{2\pi}\int_0^\infty \hbar \: \omega \: t(\omega) \frac{\partial f(\omega,T)}{\partial T}  d\omega
\label{eqn:sigma}
\end{equation}

If the exact structure of the interface is unknown one can estimate the interface conductance between material a and material b within the diffuse mismatch model \cite{Swa1989}.
For the transmission function we assume that at the interfaces the phonons lose memory of their original state and are either scattered in material a or material b. The corresponding probability is proportional to the transmission function of the bulk materials. The transmission function is normalized such that for an interface between the same material the overall transmission is 1/2 of the bulk transmission. 

\begin{equation}
t(\omega)=\frac{t_a(\omega)t_b(\omega)}{t_a(\omega)+t_b(\omega)}
\label{eqn:DDM}
\end{equation}

$t_a(\omega)$ and $t_b(\omega)$ are the bulk transmission functions.

We are using this model to estimate the influence of interfaces between two different materials. In future work we plan to investigate in addition the coherent transport across interfaces.

\section{Results and Discussion}

\begin{figure}[htbp]
  \centering
    \includegraphics[width=0.50\textwidth]{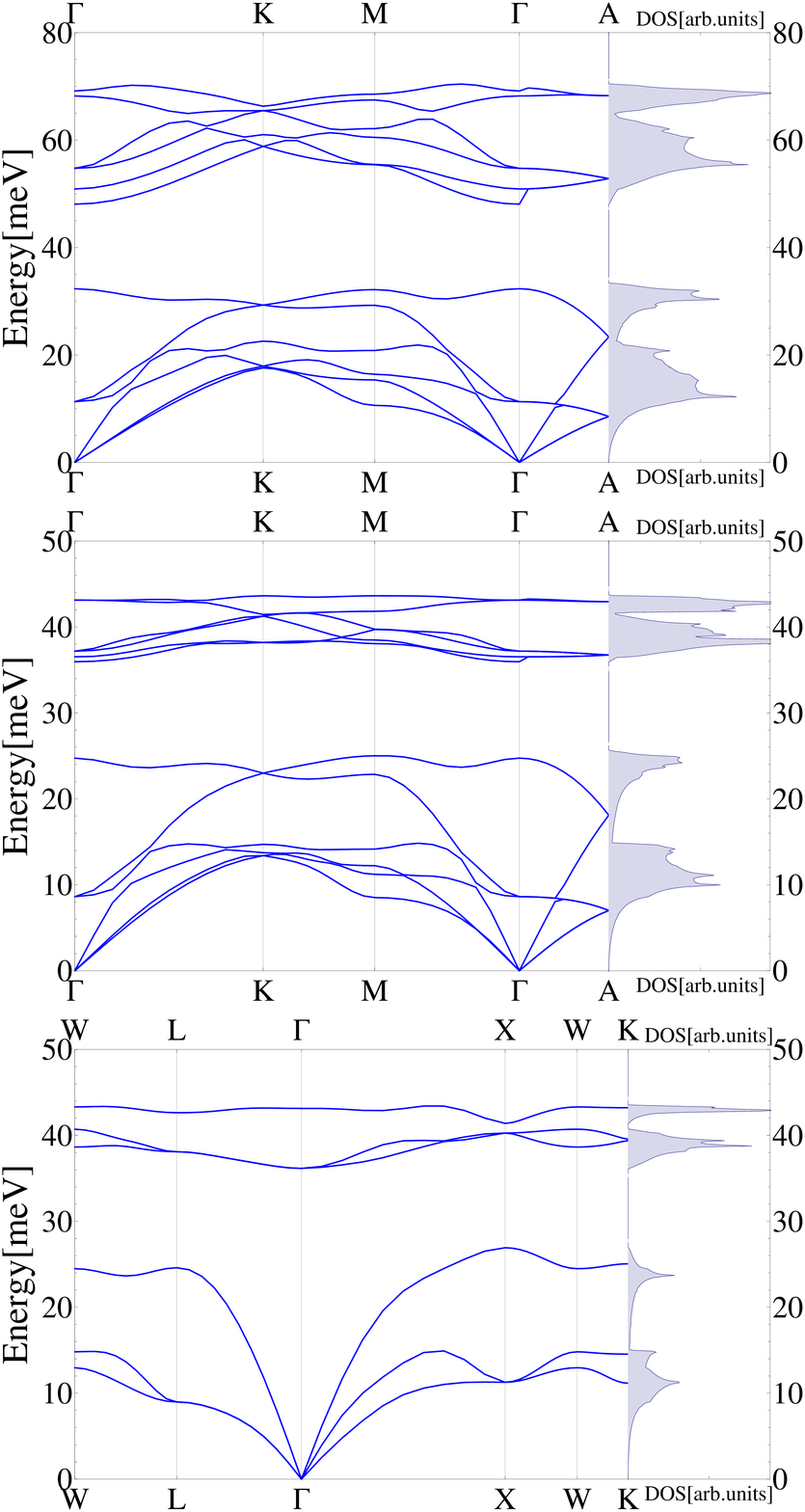} 
   \caption{Dispersion relation and density of states. Top ZnO in wurtzite structure. Center ZnS in wurtzite structure. Bottom ZnS in zincblende structure} 
  \label{fig:fig1}
\end{figure}

\begin{figure*}[htbp]
  \centering
    \includegraphics[width=0.99\textwidth]{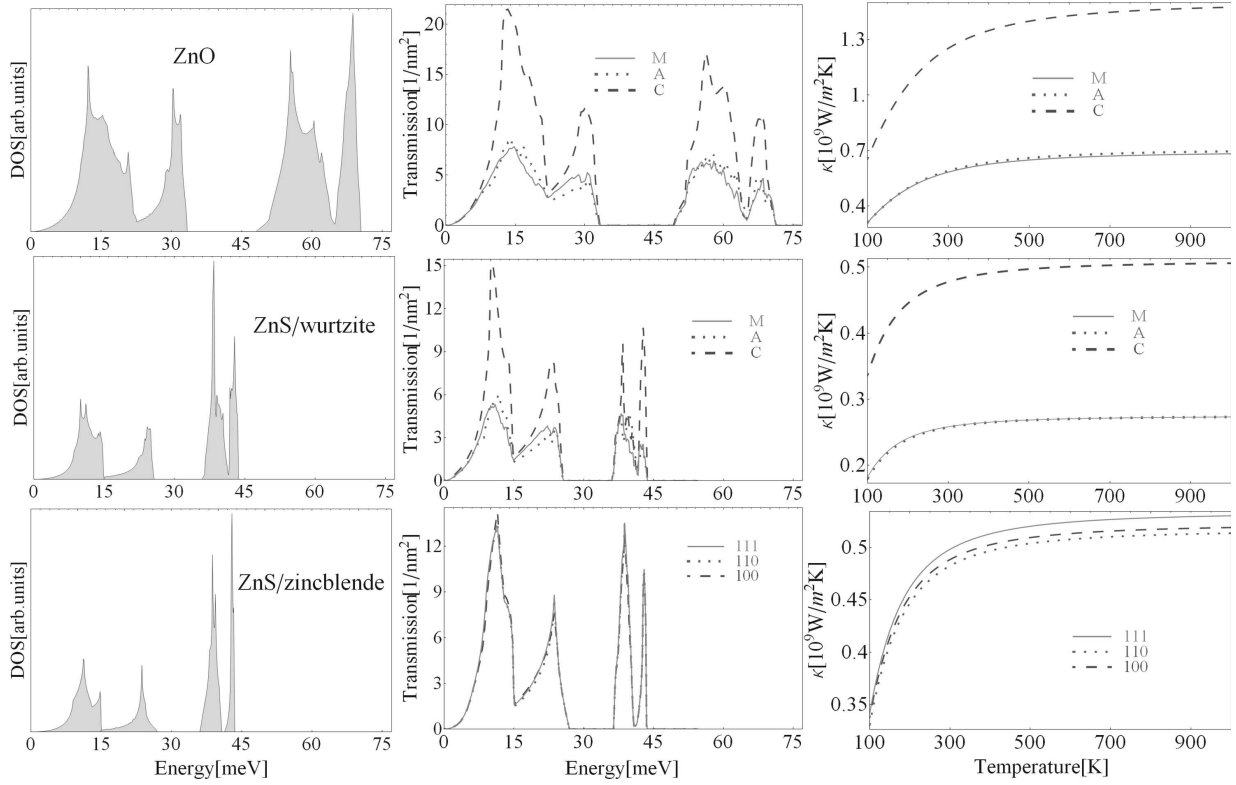} 
   \caption{Row 1: ZnO in wurtzite structure. Row 2: ZnS in wurtzite structure. Row 3: ZnS in zincblende structure. Column 1: Density of states. Column 2: Transmission function for different directions. Column 3: Temperature dependend conductance $\kappa$. C, A and M stands for the directions pependicular to the respective c-,a-,m-plane. The Miller indices for these directions are [0001], [11$\bar2$0] and [01$\bar1$0] respectively}
  \label{fig:fig2}
\end{figure*}

\textbf{Ground state calculations and cell shape optimization} are performed using a k-point sampling of 6x6x3 for ZnO wurtzite and ZnS wurtzite and a cutoff energy of 60 hartree for the plane wave expansion. For the ZnS in zincblende structure we perform ground state and cell shape optimization with a k-point grid of 6x6x6 and a cutoff energy of 60 hartree for the plane wave expansion. Using these parameters the lattice constants are converged within 0.1\%. The results of the cell shape optimization and results from other groups are shown in Table \ref{tab:tab1}. The lattice constants for ZnO wurtzite are about 1.6\% smaller than the measured ones. The volume is about 3.9\% too small. For ZnS in wurtzite structure and zincblende structure we also obtain lattice constant that are smaller than the measured one. This overestimation of the binding energies is a well known effect of the LDA. Based on the ground state density the dynamical matrix is calculated on a q-point grid in the brillouin zone. For wurtzite structure a 6x6x3 grid is used and for zincblende a 6x6x6 grid is used. The ifcs are calculated by a Fourier transform of the dynamical matrix. The ifcs are converged such that the conductance obtained from the ifcs are converged within 1\%. The dispersion relation and density of states calculated with abinit are shown in Fig.\ref{fig:fig1} for the three different material systems. The density of states and dispersion relation are in good agreement to calculations reported by other groups \cite{Cheng2009,Serrano2004}. Fig.\ref{fig:fig2} left column shows the density of states of all three material systems. It shows that the overlap between the density of states of ZnO and ZnS is rather small regardless of the structure of ZnS, which indicates a high phonon scattering at a ZnS/ZnO interface. We will emphasis this later in this paper by applying the diffuse mismatch model (DMM) to the two materials.
\begin{table*}
\centering
\begin{tabular}{p{2 cm}p{2 cm}p{2 cm}p{2 cm}p{2 cm}p{2 cm}} 
\hline
 & a(\AA)  & c(\AA) & c/a & u & V(\AA$^3$)  \\
\hline
\multicolumn{6}{c}{ZnO Wurtzite}\\ 
\hline
This Work   & 3.198   & 5.167  &  1.615 & 0.379 &  22.882  \\
Theo.\cite{Serrano2004}   & 3.198 & 5.167    &  1.615 & 0.379 &  22.882 \\
Theo.\cite{Goano2007}   & 3.23  & 5.168    &  1.6 & 0.381 &  23.3468\\
Exp.\cite{Alb1989}  & 3.2499 & 5.20658  &   1.602 & 0.3819 & 23.8119 \\
Exp.\cite{Dec2003} & 3.258 & 5.220 & 1.6022 & 0.382 &  23.9924 \\
\hline 
\multicolumn{6}{c}{ZnS Wurtzite}\\ 
This Work   & 3.755   & 6.168  &  1.643 & 0.374 &  37.659  \\
Theo.\cite{Cheng2009}   & 3.81  & 1.33    &  1.64 & - & 39.2753 \\
Exp.\cite{Kisi1989}  & 3.8227 & 6.2607   & 1.6378 & 0.3748 & 39.6154 \\
\hline 
\multicolumn{6}{c}{ZnS Zincblende}\\ 
This Work   & 5.32   &    &   &   &  37.642  \\
Theo.\cite{Cheng2009}   & 5.4  &     &   &   &  39.366 \\
Exp.\cite{Asw1960}  & 5.4109 &   &   &  & 39.6049 \\
\end{tabular}
\caption{Calculated lattice constants compared with reported experimental and theoretical values}
\label{tab:tab1}
\end{table*}

\begin{figure}[htbp]
  \centering
    \includegraphics[width=0.5\textwidth]{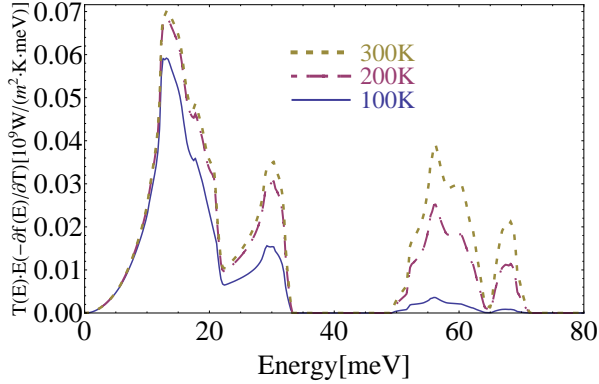} 
   \caption{Differential conductance for ZnO in C-direction at temperatures of 100 K, 200K and 300K} 
  \label{fig:fig3}
\end{figure}

\textbf{Transmission functions} for the pure materials in the different structures for different transport directions are shown in Fig.\ref{fig:fig2} middle column. The transmission functions of the wurtzite structures show a strong dependence on the transport direction. For ZnO the transmission function in c-direction is approximately a factor of 2-3 larger then for the other two directions. For ZnS the transmission function for the c-direction is also the highest. The transmission function for the zincblende structure shows no significant dependence on the direction. Using Eq.(\ref{eqn:sigma}) we calculate the conductance. In Fig.\ref{fig:fig3} the differential conductance which is in principle the integrand of Eq.(\ref{eqn:sigma}) is shown for different temperatures. It is shown that for low temperatures the main contribution of the conductance comes from phonons which have low energy. Raising the temperature leads to a small increase of the differential conductance at low energies and a strong increase at higher temperatures. At high temperatures the shape of the differential conductance is only govern by the transmission function. Now we consider the conductance of the pure ideal materials. In particular, we consider only ballistic transport thus no scattering within the material. The only scattering is due to contact resistance which is also called Sharvin resistance. Therefore, the conductance shown in Fig.\ref{fig:fig2} has the unit of an interface conductance. The conductance for the pure materials in different transport directions are shown in Fig.\ref{fig:fig2} right column. ZnO has the highest conductance in C-direction. In the other two directions ZnO has a conductance that is reduced by a factor of 3 compared to the C-direction. This behavior is already included in the transmission function of ZnO. The conductances for ZnS shows only a small difference between the different directions in zincblende structure and the C-direction in wurtzite structure. A significant smaller conductance shows ZnS in wurtzite structure in A- and M-direction. The temperature dependence of all conductances show a similar behavior. At low temperatures the conductance increases rapidly with temperature until all phonon states are sufficiently occupied. Further increase in temperature leads only to a small increase in the conductance.

\begin{figure}[htbp]
  \centering
  \begin{minipage}[b]{7.5 cm}
		\includegraphics[width=1.00\textwidth]{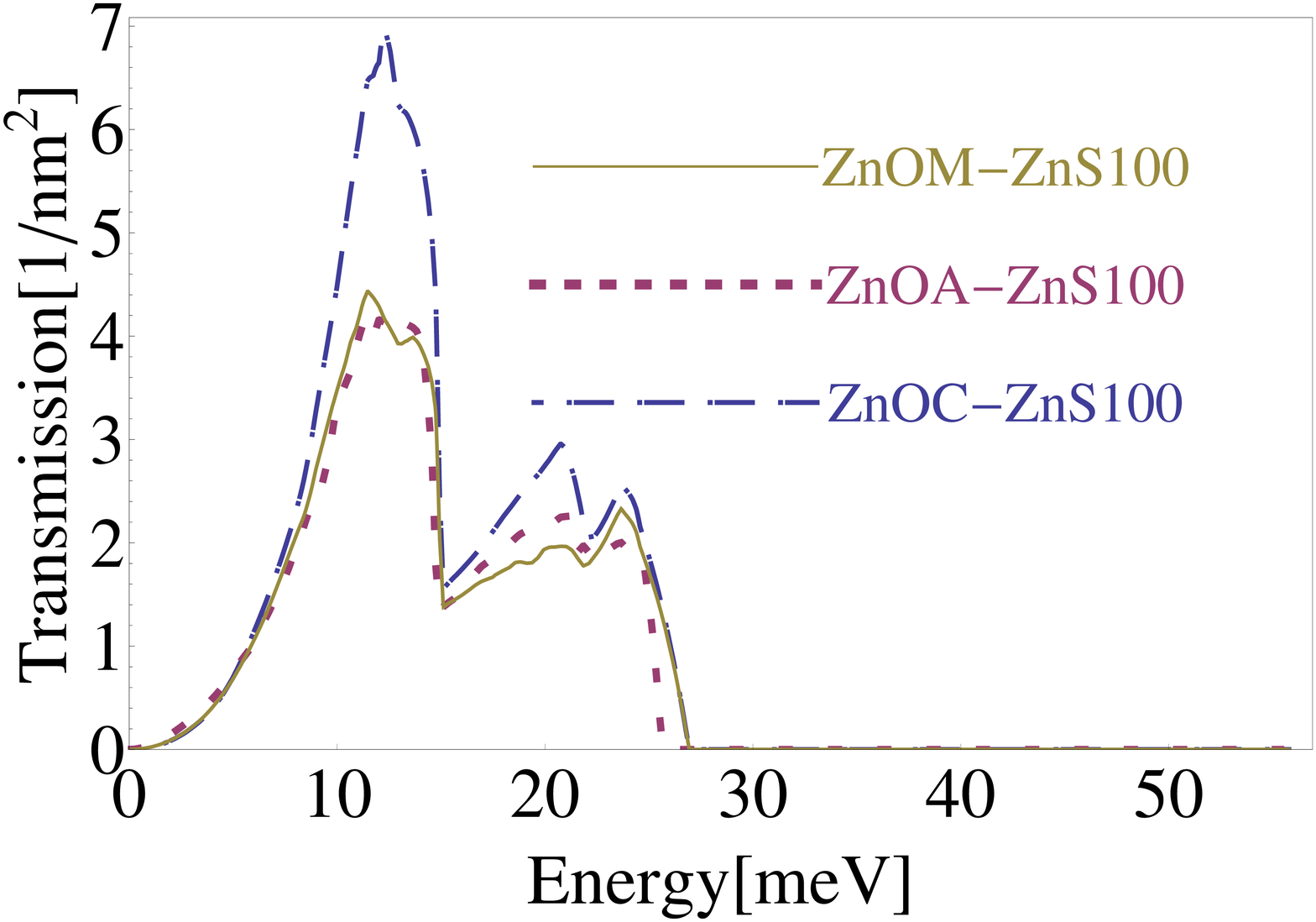}
  \end{minipage}
  \begin{minipage}[b]{7.5 cm}
		\includegraphics[width=1.00\textwidth]{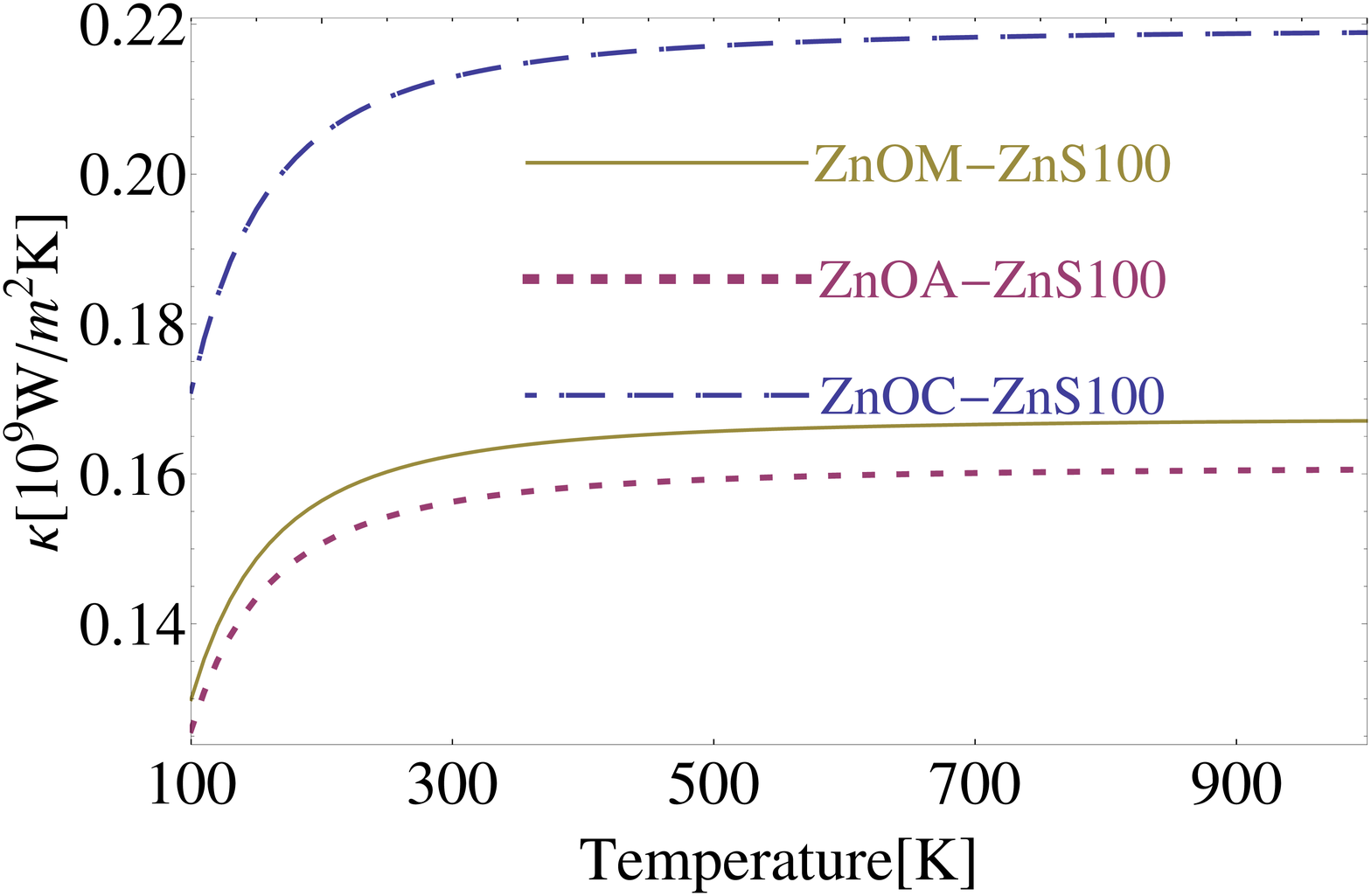}
  \end{minipage} 
   \caption{Left: Transmission function of a ZnO/ZnS100(zincblende) interfaces using diffuse mismatch model (DMM) for different orientations of ZnO. Right: Corresponding thermal conductance. ZnOC-ZnS100 stands for an interface between ZnO in C-direction and ZnS in 100 direction} 
  \label{fig:fig4}
\end{figure}

\begin{figure}[htbp]
  \centering
  \begin{minipage}[b]{7.5 cm}
		\includegraphics[width=1.00\textwidth]{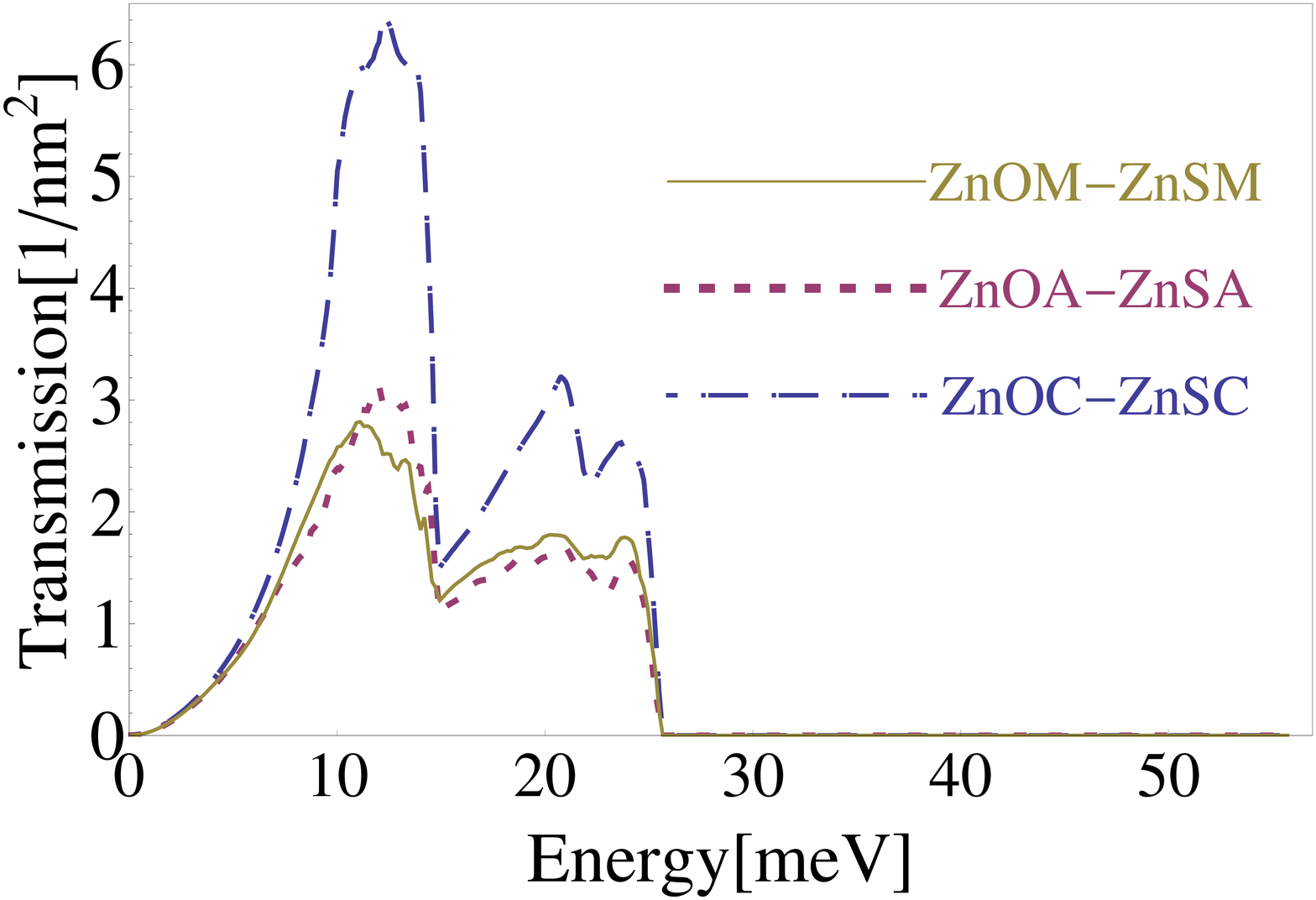}
  \end{minipage}
  \begin{minipage}[b]{7.5 cm}
		\includegraphics[width=1.00\textwidth]{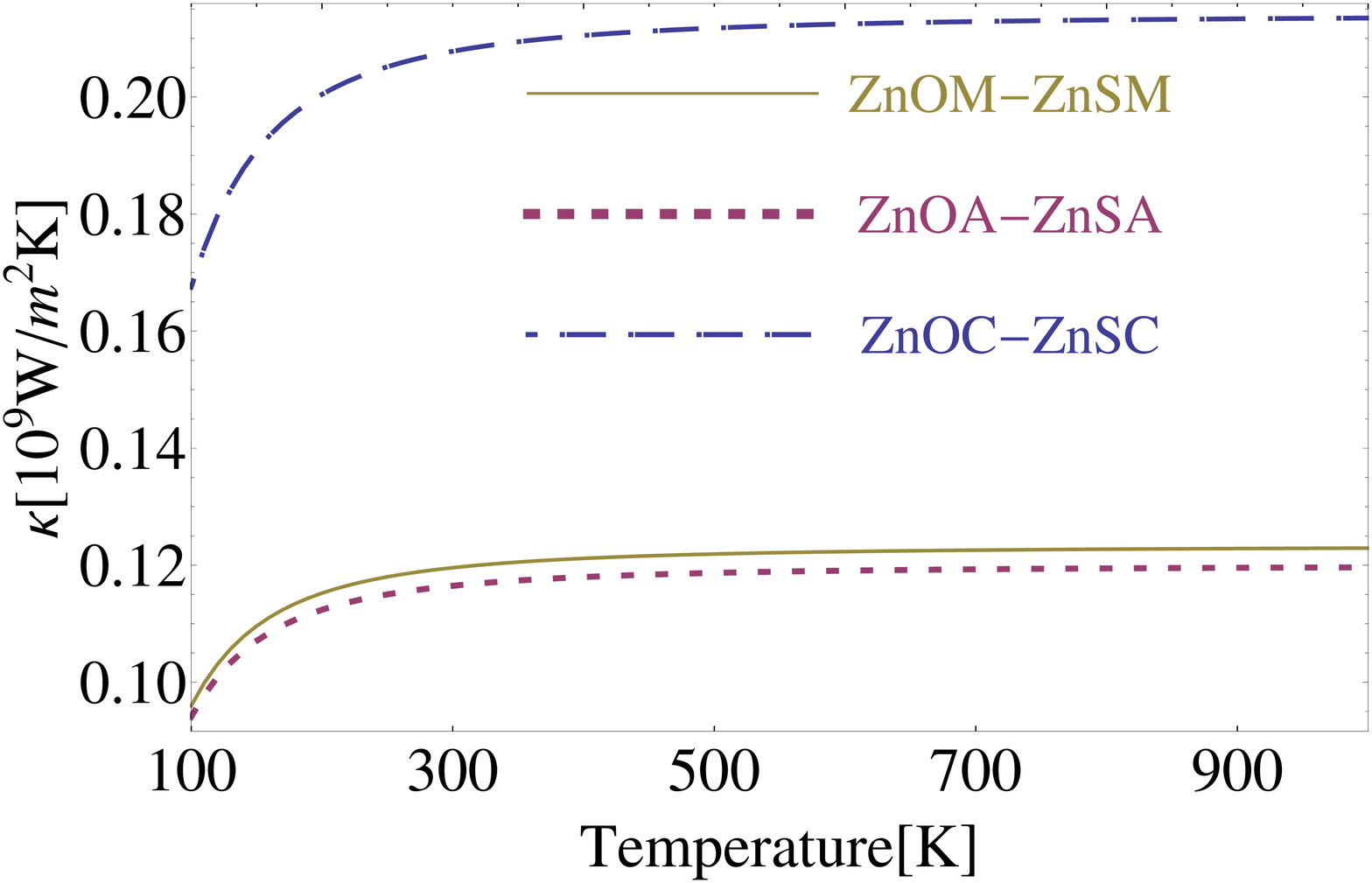}
  \end{minipage} 
   \caption{Left: Transmission function of a ZnO/ZnS(wurtzite) interfaces using DMM for different orientations of ZnO. Right: Corresponding thermal conductance} 
  \label{fig:fig5}
\end{figure}

\textbf{The diffuse mismatch model} is used to estimate the interface conductance of different combination of ZnO/ZnS interfaces. The conductance is again calculated using Eq.(\ref{eqn:sigma}). The transmission function for an interface is calculated using Eq.(\ref{eqn:DDM}). Since the transmission functions of ZnO and ZnS have no overlap above 28meV, all states above 28meV can not contribute to heat transport, which leads to a decrease in the conductance especially for higher temperatures. Since the transmission function for all directions in zincblende structure for ZnS have a similar shape we used the 100 direction representative for all directions.  Fig.\ref{fig:fig4} shows the transmission function and conductance of the interface between ZnO and ZnS in 100 direction. Additionally we investigated interfaces between ZnO and ZnS where both are in wurtzite structure. Here we focus our investigations on interfaces with the same directions. Fig.\ref{fig:fig5} shows the transmission function and conductance of the interface between ZnO and ZnS both in wurtzite structure. These interface conductances are approximately one order of magnitude higher than typical interfaces conductances reported in the literature \cite{Huang2011,Swa1987}, which is reasonable, since the DMM gives an upper estimation for the interface conductance. 
The impact of such interfaces on the figure of merit is estimated by computing the total conductance per m$^2$ of a ZnO/ZnS superlattice structure with different period length. The total conductance per m$^2$ is calculated using

\begin{equation}
G=\left[ (\frac{\kappa_{ZnO}}{L/2})^{-1}+(\frac{\kappa_{ZnS}}{L/2})^{-1}+n \cdot (\kappa_{Int})^{-1} \right]^{-1}
\label{eqn:s1}
\end{equation}
 
\begin{figure}[htbp]
  \centering
  \begin{minipage}[b]{4.4 cm}
		\includegraphics[width=1.00\textwidth]{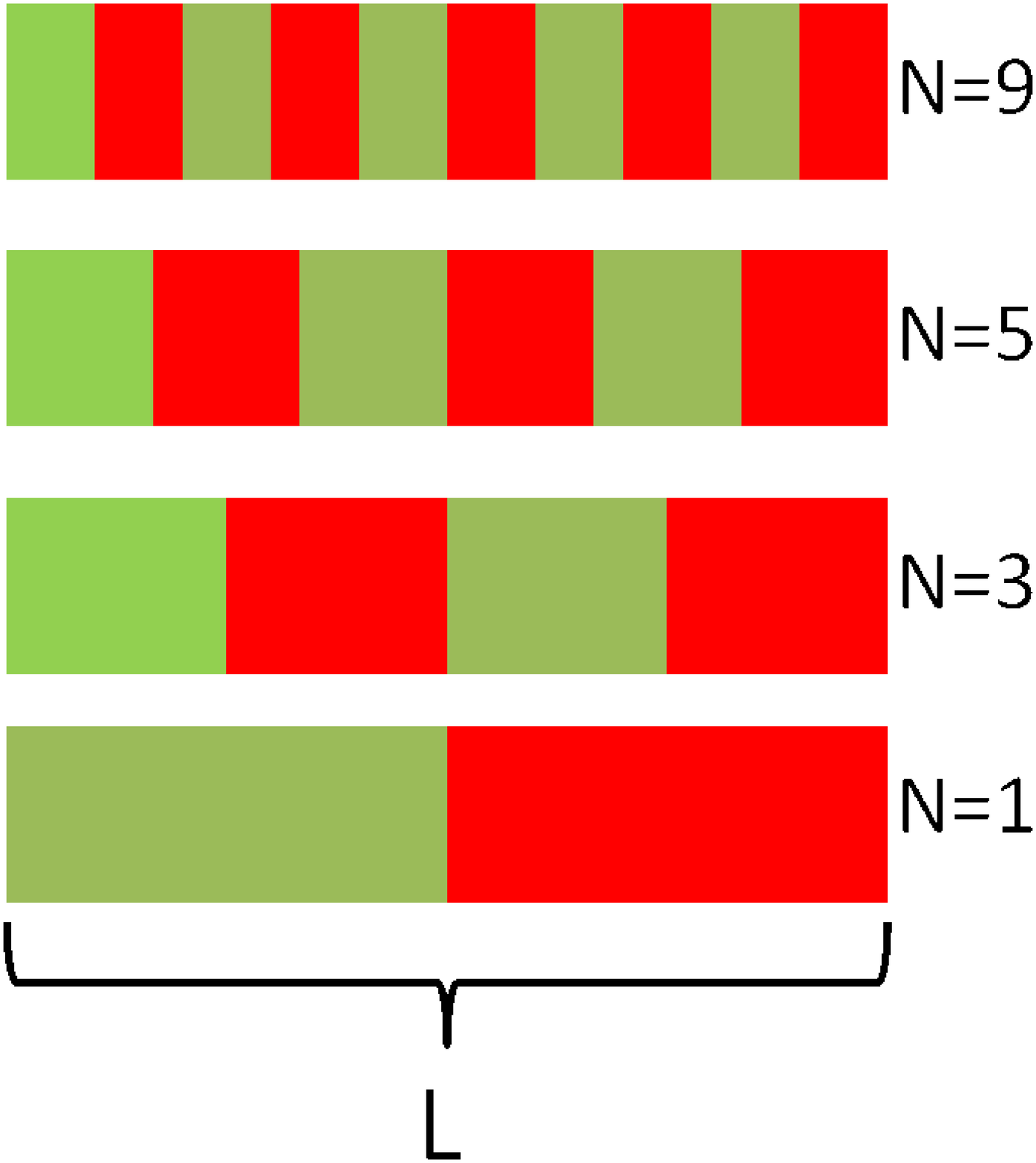}
  \end{minipage}
  \begin{minipage}[b]{7.6 cm}
		\includegraphics[width=1.00\textwidth]{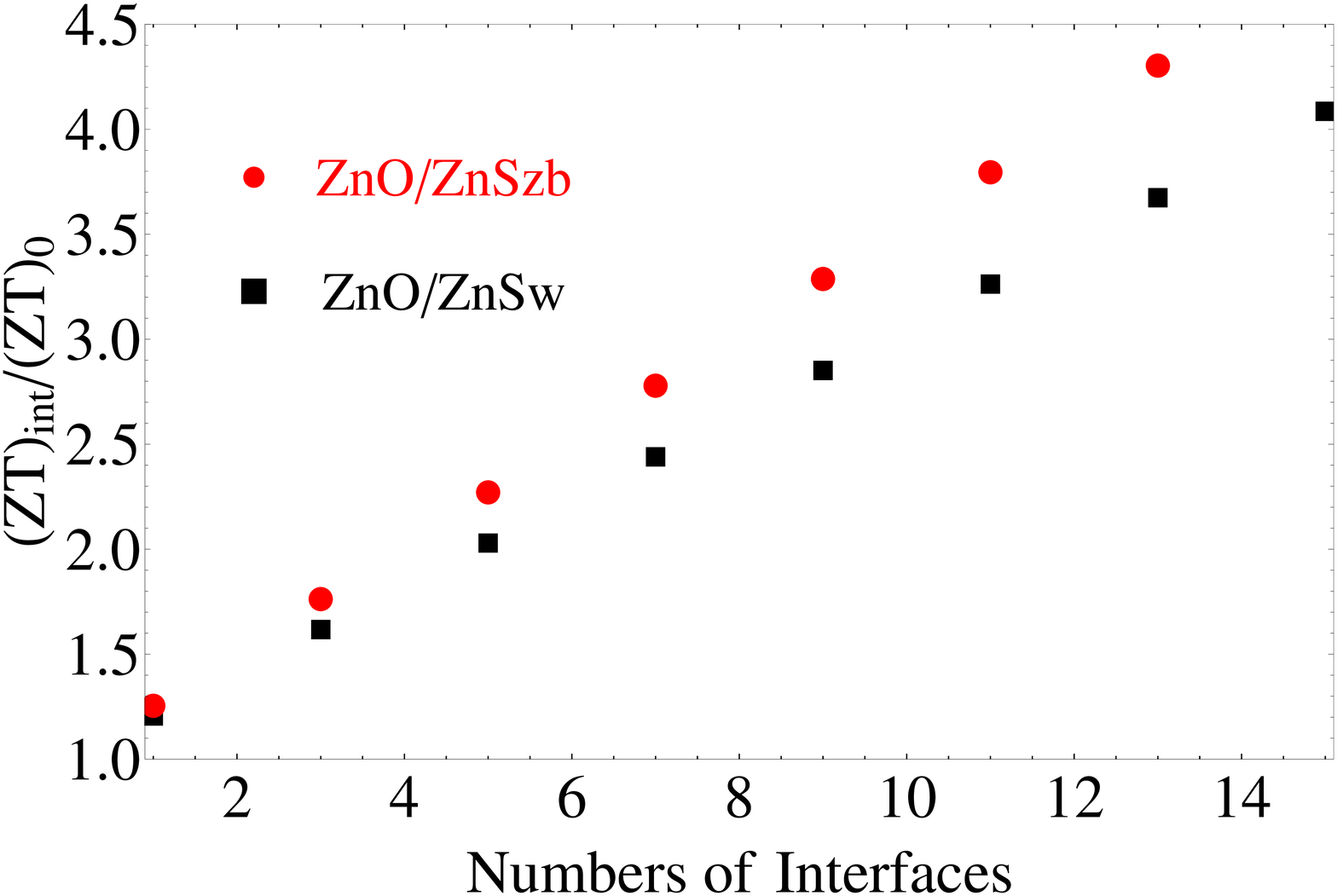}
  \end{minipage} 
   \caption{Left: Sketch of a ZnO/ZnS superlattice structure of length L. Right: Figure of merit as a function of the number of interfaces within the superlattice structure. The used parameter are given in Table \ref{tab:tab2}} 
  \label{fig:fig6}
\end{figure}

L is the length of the structure and n is the number of interfaces. $\kappa_{ZnO}$ and $\kappa_{ZnS}$ are the bulk conductivities and $\kappa_{Int}$ is the interface conductance. Fig.\ref{fig:fig6} shows the figure of merit with different interfaces divided by the figure of merit without interfaces. In this calculation we assume that the interfaces have only an impact on the phonon thermal conductance and that the phonon thermal conductance is much larger than the electric part. Thus the ratio of $(ZT)_{int}$ with interfaces and by ZT without interfaces is given by $\kappa_{ZT}/\kappa_{ZT_{int}}$. For $\sigma_{Int}$ we used the average value of the interface conductance of ZnO/ZnS/wurtzite and ZnO/ZnS/zincblende presented in this paper. All parameters we used for this plot are listed in Table \ref{tab:tab2}. 
It can be seen, that the figure of merit increasing linearly with the number of interfaces. The slope of this curve is given by the inverse of the interface conductance. Consequently superlattice structures of ZnO/ZnS are promising for a substantial increase of the figure of merit in thermoelectric devices.
\begin{table}
\centering
\begin{tabular}{p{2 cm}|p{5 cm}} 
Parameter  & Value    \\
\hline
$\kappa_{ZnO}$\cite{Slack1972}  & 54W/mK    \\
$\kappa_{ZnS}$\cite{Slack1972}  & 27W/mK   \\
$\kappa_{ZnO/ZnSw}$  & 0.175$\cdot 10^{9}$ W/m$^2$K\\
$\kappa_{ZnO/ZnSzb}$  & 0.14$\cdot 10^{9}$ W/m$^2$K\\
L & $1.0\cdot 10^{-6}$m \\
\end{tabular}
\caption{Parameters and references used for Fig. \ref{fig:fig6}. The interface conductance $\kappa_{ZnO/ZnSw}$ and $\kappa_{ZnO/ZnSzb}$ are the average values of the different interfaces shown in Fig.\ref{fig:fig5} and Fig.\ref{fig:fig4}. The values are taken at a temperature of 300K. $\kappa_{ZnO}$ is the ZnO bulk thermal conductivity at 300K average over the different directions \cite{Slack1972}. $\kappa_{ZnS}$ is the ZnS zincblende bulk thermal conductivity at 300K average over the different directions \cite{Slack1972}.}
\label{tab:tab2}
\end{table}

\section{Summary and Conclusion}
We performed DFT calculations of ZnO in wurtzite structure and ZnS in zincblende and wurtzite structure in the LDA. We calculated the lattice constants for this structures which agree well with other reported values. From DFT calculations we also obtain phonon band structures and phonon density of states which also agree well with reported ones by other groups. Based on these calculations we can also obtain the interatomic force constants (ifcs). These ifcs were then used within an AGF-Method to calculate bulk transmission functions. The difuse missmatch model (DMM) is used to estimate the interface conductance between ZnS and ZnO. We observed that the existence of such interfaces can have a significant impact on the thermal conductance and could lead to a substantial increase in the figure of merit. The DMM is only a rough method to estimate the interface conductance. In order to improve the description of the interface one can use the AGF method to built layer structures and calculate the transmission function of an interface directly from Eq.(\ref{eqn:Transmission}), which we plan in the future.

\section{Appendix}
\begin{figure}[htbp]
  \centering
    \includegraphics[width=0.50\textwidth]{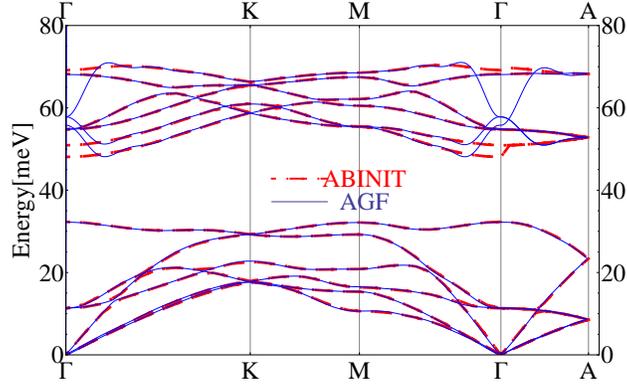} 
   \caption{Dispersion relation calculated with ABINIT (dashed) and the AGF method (solid)} 
  \label{fig:fig7}
\end{figure}

\begin{figure}[htbp]
  \centering
    \includegraphics[width=0.50\textwidth]{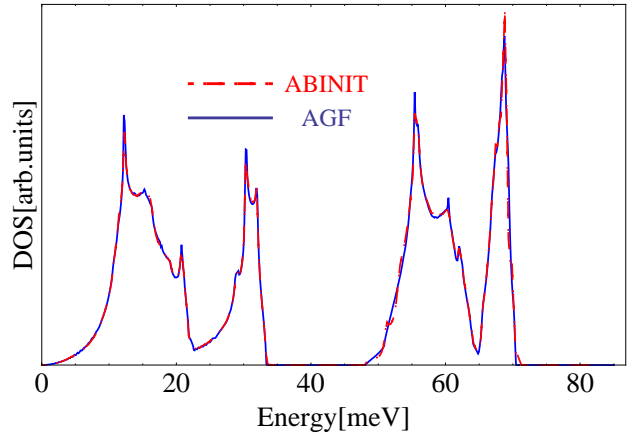} 
   \caption{Desity of states calculated with ABINIT (dashed) and the AGF method (solid)} 
  \label{fig:fig8}
\end{figure}

\begin{figure}[htbp]
  \centering
    \includegraphics[width=0.50\textwidth]{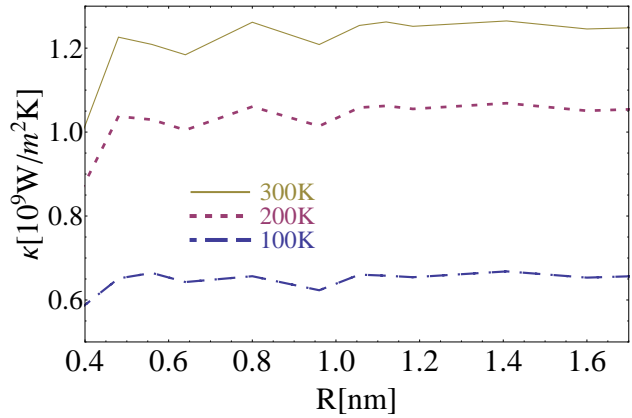} 
   \caption{Conductance of ZnO in C-direction calculated as a function cutoff radii at different temperatures} 
  \label{fig:fig9}
\end{figure}

ZnO and ZnS are both intrinsic insulators which have non-vanishing effective charges, which means that the ifcs exhibits a dipol-dipol interaction. These dipol-dipol interaction shows an $1/r^3$ decrease in real space which makes it hard to converge the calculation by summing up all interactions in real space. To avoid this problem ABINIT decomposed the ifcs in a short range part and a dipol-dipol part. The dipol-dipol part is calculated in reciprocal space (for details see \cite{Gonze1997b}). Unfortunately, in the AGF method the ifcs has to be defined in real space and therefore only a finite number of interactions can be taken into account. To estimate the error caused by this, we compared the dispersion relation and density of calculated with ABINIT and the dispersion relation and DOS calculate with the AGF. The DOS can be calculated in the AGF by using Eq.(\ref{eqn:GDos3}). Fig.\ref{fig:fig7} shows the dispersion relation calculated with the different methods. It can be seen, that both are almost equal except of three optical phonon modes which deviate in the nearness of the $\Gamma$-point. Fig.\ref{fig:fig8} shows the DOS calculated with the different methods and it can be seen that there is only a very small difference due to the mentioned three optical phonon modes. Since there are only three optical phonon modes which are described wrong only around the $\Gamma$-point we can conclude that the error we made by truncating the interactions has only a small effect on total conductance. To estimate the radius after which we can truncate the sum of the interaction we perform convergence tests with respect to the transmission function.
In Fig.\ref{fig:fig9} the conductance for ZnO in C-direction is calculated using different radii for the interactions. After a radius of 1.12nm the conductance is convergend within 1\%. The visible very small oscillations for large cutoff radii are due to the mentioned wrongly describes optical modes close to the $\Gamma$-Point. In this paper we use a cutoff radius of at least 1.12nm.

{\sl Acknowledgements} We thank the Deutsche Forschungsgemeinschaft for supporting us in the framework of the priority programme 1386 `Nanostructured thermoelectrics'.

\bibliographystyle{unsrt}
\bibliography{ref}

\end{document}